\documentclass[twocolumn,nopacs,preprintnumbers,amsmath,amssymb]{revtex4}

\usepackage{verbatim}
\usepackage{graphicx}% Include figure files
\usepackage{dcolumn}% Align table columns on decimal point
\usepackage{bm}% bold math

\begin{document}

\title{Self-aligned carbon nanotube transistors with charge transfer doping}
\author{Jia Chen}
\email{chenjia@us.ibm.com}
\author{Christian Klinke}
\author{Ali Afzali}
\author{Phaedon Avouris}
\email{avouris@us.ibm.com}
\affiliation{IBM Research Division, T.
J. Watson Research Center, PO Box 218, Yorktown Heights, New York
10598}

\begin{abstract}

This letter reports a charge transfer \textit{p}-doping scheme
which utilizes one-electron oxidizing molecules to obtain stable,
unipolar carbon nanotube transistors with a self-aligned gate
structure. This doping scheme allows one to improve carrier
injection, tune the threshold voltage \textit{V}$_{\mathit{th}}$,
and enhance the device performance in both the ``ON-'' and
``OFF-'' transistor states. Specifically, the nanotube transistor
is converted from ambipolar to unipolar, the device drive current
is increased by 2\ensuremath{\sim}3 orders of magnitude, the
device OFF current is suppressed and an excellent
\textit{I}$_{\mathit{ON}}$\textit{/I}$_{\mathit{OFF}}$ ratio of
10$^{6}$ is obtained.  The important role played by metal-nanotube
contacts modification through charge transfer is demonstrated.

\end{abstract}

\maketitle

Significant progress has been made recently on carbon nanotube
(CNT) based field effect transistors (FET), both in terms of their
fabrication and the understanding of their performance
limits.$^{1,2,3}$ Nevertheless, there are still key issues to be
addressed. In particular, there has been no fabrication
process-compatible doping method for CNTs. Unlike doping of
metal-oxide-semiconductor (MOS) FETs, CNTFETs cannot be doped
substitutionally via ion implantation which destroys the CNT
lattice.   Instead, the quasi one-dimensional (1D) CNTFETs can be
doped via charge transfer processes.   Benefiting from the weak
screening, doping of 1D structures can be more efficient than in
3D devices (e.g., in MOSFETs).$^{4}$ Current methods used to
obtain \textit{p}- or \textit{n}- CNTFETs suffer from severe
process limitations:  for example, a \textit{p}-CNTFET can be
converted to an ambipolar or an \textit{n}-FET under vacuum
pumping through the change of the oxygen content at the metal-CNT
interface.$^{5}$  Efforts utilizing exposure to gas-phase
NO$_{2}$$^{6}$ or trifluoro-acetic acid (TFA)$^{7}$ to obtain
\textit{p}-CNTFETs require a controlled environment to prevent
dopant desorption, and the devices degrade quickly upon exposure
to air.  Performance-wise, the Schottky barriers (SB) formed
between the CNT and the source/drain metal contacts lead to
contacts-dominated switching, $^{8}$ to a large subthreshold swing
\textit{S}=\textit{dV}$_{\mathit{gs}}$\textit{/d(log}
\textit{I}$_{\mathit{d}}$\textit{)}, where
\textit{V}$_{\mathit{gs}}$ is the gate voltage and
\textit{I}$_{\mathit{d}}$ is the drain current,  and to strong
ambipolar conduction when the transistor is scaled down
vertically, i.e., when the gate insulator is thinned
substantially.  The resulting SB-limited drive current
(\textit{I}$_{\mathit{ON}}$), the slow switching, and
exponentially increasing OFF current (\textit{I}$_{\mathit{OFF}}$)
with increasing drain field$^{9}$ are unacceptable in logic gate
applications.  To make CNTFETs technologically viable, it is
therefore crucial to find doping methods and materials that both
are fabrication process stable and meet the performance
challenges. Here we report on a chemical \textit{p}-doping scheme
utilizing oxidizing molecules and a charge transfer mechanism to
obtain self-aligned, air-stable, unipolar CNTFETs. We demonstrate
the ability to change carrier injection properties; to tune
threshold voltage (\textit{V}$_{\mathit{th}}$), and to improve
device performance in both ON- and OFF-states.

We have fabricated CNTFETs using laser-ablation-produced CNTs,
palladium metal source and drain electrodes separated by 300 nm on
top of 10 nm SiO$_{2}$, and a Si backgate. The substrate with the
CNTFETs were then immersed in a dilute solution (0.01
\ensuremath{\sim} 0.1 mg/mL) of triethyloxonium
hexachloroantimonate (C$_{2}$H$_{5}$)$_{3}$O$^{+}$SbCl$_{6}$$^{-}$
(\textbf{OA}) in methylene chloride or dichlorobenzene.
\textbf{OA} is a one-electron oxidant which is known to oxidize
aromatic compounds and form charge transfer complexes.$^{10}$ The
interaction of a CNT with this molecule generates a positively
charged CNT stabilized by the negatively charged counter ion
SbCl$_{6}$$^{-}$. If \textbf{\ensuremath{\alpha}} is used to
represent the six carbon atom (benzene-like) ring on a CNT, the
interaction between the CNT and \textbf{OA} can be represented by
Eq. (1):

2 \textbf{\ensuremath{\alpha}} \textbf{+} 3
\textbf{[}(C$_{2}$H$_{5}$)$_{3}$O$^{+}$SbCl$_{6}$$^{-}$]
\ensuremath{\rightarrow} 2 [\textbf{\ensuremath{\alpha}}$^{+.}$
SbCl$_{6}$$^{-}$ ] + 3 C$_{2}$H$_{5}$Cl + 3
(C$_{2}$H$_{5}$)$_{2}$O + SbCl$_{3}$  \verb"  " (1)

Among the reaction products, C$_{2}$H$_{5}$Cl and
(C$_{2}$H$_{5}$)$_{2}$O are volatile, while SbCl$_{3}$ and excess
dopants are removed by rinsing with solvents. Typical transistor
transfer characteristic curves (\textit{I}$_{\mathit{d}}$
vs\textit{. V}$_{\mathit{gs}}$) at drain bias
(\textit{V}$_{\mathit{ds}}$) = -0.1 to -0.5 V at -0.1 V step of a
CNTFET before and after doping are shown in Figs. 1(a) and 1(b),
respectively. We find (i) \textit{V}$_{\mathit{th}}$ for hole
conduction increased from -0.7 V (intrinsic CNTFET) to 0.1 V; (ii)
the drive current \textit{I}$_{\mathit{ON}}$ improved by 2 orders
of magnitude, greatly reducing the contact resistance between tube
and metal; (iii) \textit{S} decreased from 200 to 85 mV/decade,
indicating improved switching and a strong reduction of the
Schottky barrier; (iv) the transistor transfer characteristic
changed from the original ambipolar to pure \textit{p}-type
conduction. After doping, we successfully suppressed the minority
carrier (in this case, the electron current) injection at the
drain, and obtained excellent drain-induced barrier lowering
(DIBL) behavior as is clearly shown from Fig. 1(b). The above
results were reproducible with 20 different CNTFET devices stored
under nitrogen for weeks. These devices preserve their doping
characteristics after solvent washing and vacuum pumping.

\begin{figure}[ht]
\begin{center}
\includegraphics[width=0.45\textwidth]{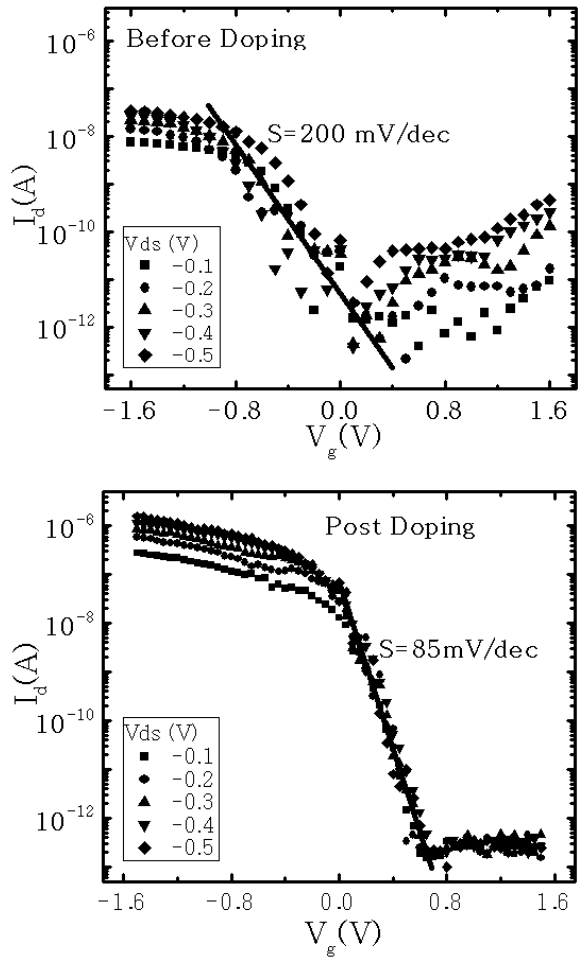}
\caption {\it Transfer characteristics of a CNFET before (a) and
after (b) \textbf{OA}-doping at \textit{V}$_{\mathit{ds}}$ = -0.1
to -0.5 V @ step of -0.1 V.}
\end{center}
\end{figure}

\begin{figure}[ht]
\begin{center}
\includegraphics[width=0.45\textwidth]{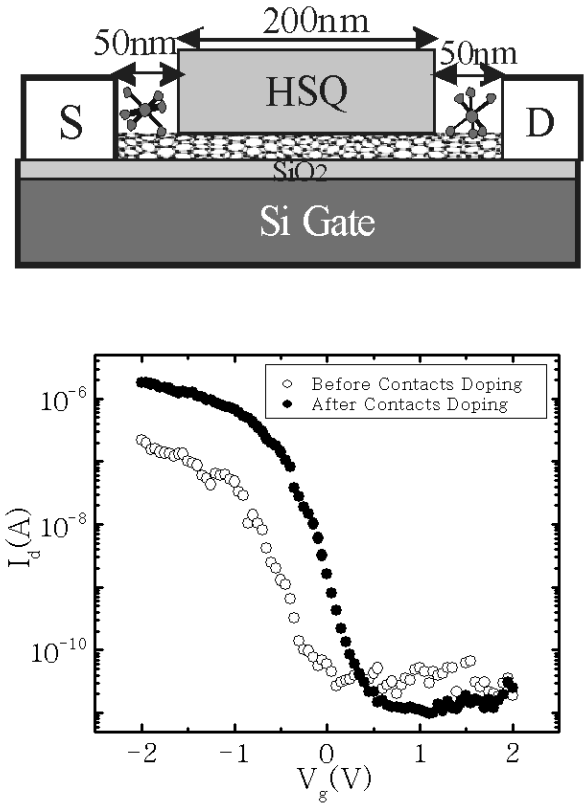}
\caption {\it (a) Cross-section schematics of a HSQ patterned
CNFET selectively doped by \textbf{OA} at the source-drain
contacts; (b) transfer characteristics of the device before and
after selective \textbf{OA} doping at \textit{V}$_{\mathit{ds}}$ =
-0.5 V.}
\end{center}
\end{figure}

To further understand the dopant-CNT interaction, we selectively
doped CNTFETs at the vicinity of the contacts. The negative
electron-beam resist hydrogen silsesquioxane (HSQ), which is
resistant to the action of the solvents used for doping, was used
to mask 200 nm of the CNT channel, leaving a 50 nm unmasked gap at
each contact. These open areas were then chemically doped. A
cross-section schematic of the device is shown in Fig 2(a). The
device transfer characteristics before and after doping are shown
in Fig. 2(b).  An order of magnitude increase in the drive current
is observed, as well as an improvement in \textit{S} and a 0.5 V
increase of \textit{V}$_{\mathit{th}}$. The comparable performance
improvements in the case of contacts doping and full-device doping
discussed earlier point out the crucial role the contacts play in
a SB CNTFET. The simultaneous increase of
\textit{I}$_{\mathit{on}}$, decrease of \textit{S} and the
suppression of minority carrier injection [Fig. 1(b)] cannot be
explained simply by a carrier concentration increase, as in the
case of doping in a conventional MOSFET, where substitutional
dopants modify the Fermi level in the bulk of the channel.  The
improvement in both the device ON- and OFF- states after doping of
a CNTFET can, however, be understood in terms of its SB height
reduction. The contact barriers were manipulated via the work
function modification of the source and drain electrodes under
chemical doping. When the \textbf{OA} molecules interact with the
electrodes, charge transfer takes place and the stable ions,
SbCl$_{6}$$^{-}$, remain at the surface of the metal electrodes.
They generate an outwards-directed surface dipole that reinforces
the intrinsic surface dipole, thus raising the metal workfunction.
An increased work function favors hole injection at one electrode
(high \textit{I}$_{\mathit{on}}$ and low \textit{S}), while it
suppresses electron injection at the other electrode. In fact,
modification at the metal-CNT interface band lineup (SB), due to
local variation of substrate work function induced by oxygen
adsorption, has been observed directly in electrostatic force
microscopy (EFM) and scanning Kelvin probe microscopy
(SKPM).$^{11}$ It has also been shown that chemical coadsorption
on CNTFETs leads to modified SBs$^{7,}$$^{11}$ and can convert a
\textit{p}-type to an ambipolar CNTFET.$^{11}$ After doping, the
device transconductance \textit{g}$_{\mathit{m}}$
\textit{(dI}$_{\mathit{d}}$\textit{/dV}$_{\mathit{g}}$\textit{)}
at \textit{V}$_{\mathit{ds}}$ \textit{=} \textit{-1.32} \textit{V}
is \textit{2} \textit{\ensuremath{\mu}}\textit{S}, outperforming
those of the best \textit{p}-CNFET with small diameter CNTs and
the same effective gate dielectrics used in our work.

\begin{figure}[ht]
\begin{center}
\includegraphics[width=0.45\textwidth]{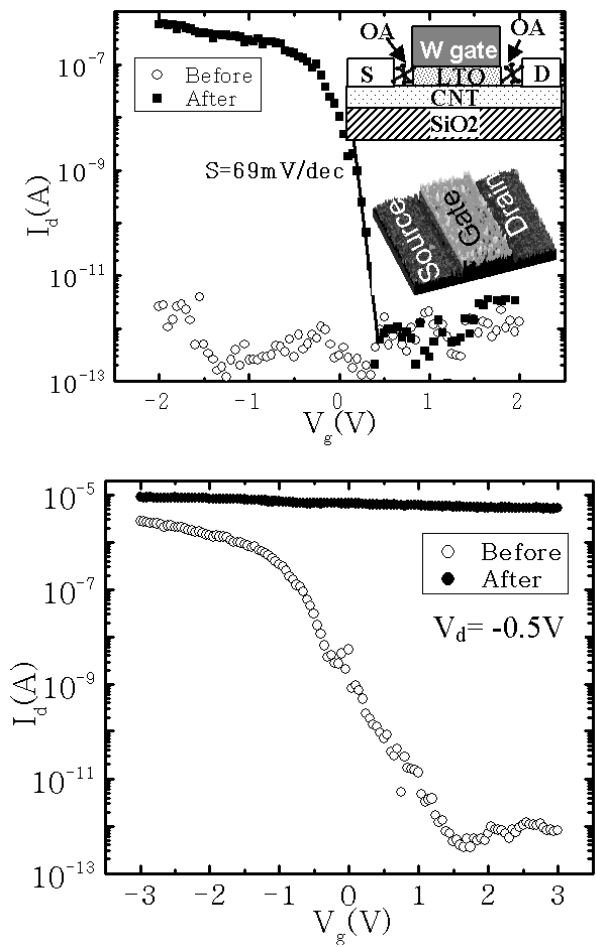}
\caption {\it (a) Transfer characteristics of a W-top-gated CNTFET
before and after \textbf{OA} ``extension doping'' at
\textit{V}$_{\mathit{ds}}$ = -0.5 V. \textit{S} is 69 mV/dec post
doping. The inset shows a schematics and a SEM picture of the
device; (b) transfer characteristics of a bottom-gated CNTFET
device before and after degenerate \textbf{OA} doping where a
\textit{p}-CNFET was transformed into an almost metallic tube.}
\end{center}
\end{figure}

We have demonstrated the reduction of SBs through charge transfer.
Ultimately, we need to minimize SBs to the point that transport
through a CNTFET is not limited by the contacts; rather by the
channel of the tube, as in a conventional MOSFET.  In a MOSFET,
the gate controls the channel region to overcome its thermionic
barrier; whereas in a SB CNTFET, strong gate coupling is crucial
to modulate the barriers at the contact regions. In the following
experiment, we utilize charge transfer to replace the gate control
over the contact regions in a CNTFET, thereby transforming it from
a contacts-dominated SB-FET to a channel-controlled FET. To
realize this, we exposed part of the channel that is in contact
with the metal electrodes to chemical dopants, and used a top gate
to individually address each device. A schematic of the device
structure is shown in the inset of Fig. 3(a). An initially
bottom-gated CNTFET was covered with 12 nm of low temperature
oxide (LTO)$^{1}$$^{2}$ and a tungsten (W) top gate was defined by
electron-beam lithography and liftoff. The source/drain separation
of the FET was 2 \ensuremath{\mu}m and the top-gated region was
1.66 \ensuremath{\mu}m. The patterned W gate was then used as an
etch mask to define the 170 nm wide chemical doping windows,
formed by removing the LTO with diluted HF, adjacent to the metal
contacts. The above process aligns the doping windows with the
top-gate in a single patterning step, resembling the ``extension
doping'' of a typical MOSFET.$^{13}$ This ``self-aligned'' process
allows one to minimize the gate to source/drain capacitive
coupling and form completely symmetric source and drain electrodes
in a FET, which is crucial in logic applications.  To reduce the
electrostatic coupling from the bottom gate, the top-gated devices
were fabricated on a 100 nm-thick oxide substrate. We do not
expect that a SB CNTFET with the above long channel configuration
can be either turned on by the weakly coupled top gate or the
source/drain fringe fields.$^{1}$$^{4}$ Indeed, the device
remained in its OFF state through a range of applied gate biases,
as shown by the open circles in Fig. 3(a). The device cannot be
turned ON unless the contact barriers are overcome and enough
carriers are induced on the ungated segments. This can be realized
by either chemical or electrostatic doping.$^{5,15,16,17}$ In this
work, we adopt the degenerate chemical doping approach to link the
source/drain contacts and the bulk of the CNT.  Figure 3(b) shows
a conventional bottom-gated CNTFET after degenerate doping with
\textbf{OA}. The \textit{p}-CNTFET was converted to an almost
metallic CNT which is not modulated by the gate bias.  When the
device was degenerately doped, carrier density in the CNT was
significantly enhanced in addition to the modification of the
contacts. Applying this degenerate doping to the top-gated device
in Fig. 3(a), we were able to successfully switch it ON and OFF,
as shown by the filled circles. After doping, \textit{p}-type
characteristics with a \textit{V}$_{\mathit{th}}$ of -0.2 V, a
sharp \textit{S} of 69 mV/dec, and an excellent
\textit{I}$_{\mathit{on}}$\textit{/I}$_{\mathit{OFF}}$ ratio of
10$^{6}$ were obtained.  Compared with the fully doped
bottom-gated device (Fig. 1), or the the much reduced parasitic
capacitance between source/drain and gate in the top gated device
(Fig. 3) yields a smaller \textit{S}, therefore faster transistor
switching.

In conclusion, we have demonstrated stable chemical
\textit{p}-doping of CNTFETs, via charge transfer: we improved
\textit{I}$_{\mathit{ON}}$ by 2\ensuremath{\sim}3 orders of
magnitude, suppressed minority carrier injection and achieved
immunity from drain induced \textit{I}$_{\mathit{OFF}}$
degradation in intrinsic SB CNTFETs,  in addition, an excellent
\textit{I}$_{\mathit{ON}}$\textit{/I}$_{\mathit{OFF}}$ ratio of
10$^{6}$, good DIBL behavior, and a tunable
\textit{V}$_{\mathit{th}}$ were obtained.  Through selective
doping we determined that most of the device characteristics
improvement came from the dopants interacting with the metal-CNT
contacts.  Finally, we demonstrated degenerate chemical doping and
realized a self-aligned top-gated CNTFET.

The authors thank K. Chan for LTO film deposition, P. Solomon and
J. Appenzeller for valuable discussions, and B. Ek for expert
technical assistance.  C. Klinke acknowledges gratefully the Swiss
National Science Foundation (SNF) for their financial support.

% \newpage
\begin{center}
\line(1,0){100.0}
\end{center}

$^{1}$ P. Avouris, MRS Bull. \textbf{29}\textbf{,} 403 (2004).

$^{2}$ P. L. McEuen, M. S. Fuhrer, and H. Park, IEEE Trans.
Nanotechnol. \textbf{1}\textbf{,} 78 (2002).

$^{3}$ A. Javey, Q. Wang, W. Kim, and H. Dai, IEDM Tech. Digest,
2003, p. 31.

$^{4}$ F. Leonard and J. Tersoff, Phys. Rev. Lett. \textbf{83}, 5174 (1999).

$^{5}$ V. Derycke, R. Martel, J. Appenzeller, and Ph. Avouris,
Appl. Phys. Lett. \textbf{80}, 2773 (2002).

$^{6}$ J. Kong, N. R. Franklin, C. Zhou, M. G. Chapline, S. Peng,
K. Cho, and H. Dai, Science \textbf{287}, 622 (2000).

$^{7}$ S. Auvray, J. Borghetti, M. F. Goffman, A. Filoramo, V.
Derycke, J. P. Bourgoin, and O. Jost, Appl. Phys. Lett.
\textbf{84}, 5106 (2004).

$^{8}$ J. Appenzeller, J. Knoch, V. Derycke, R. Martel, S. Wind,
and Ph. Avouris, Phys. Rev. Lett\textit{.} \textbf{89}, 126801
(2002).

$^{9}$ M. Radosavljevic, S. Heinze, J. Tersoff, and Ph. Avouris,
Appl. Phys. Lett. \textbf{83}, 2435 (2003).

$^{10}$ R. Rathore, A. S. Kumar, S. V. Lindeman, and J. K. Kochi,
J. Org. Chem. \textbf{63}, 5847 (1998).

$^{11}$ X. Cui, M. Freitag, R. Martel, L. Brus, and Ph. Avouris,
Nano Lett. \textbf{3}, 783 (2003).

$^{12}$ S. J. Wind, J. Appenzeller, R. Martell, V. Derycke, and
Ph. Avouris, Appl. Phys. Lett. \textbf{80}, 3817 (2002).

$^{13}$ S. Wolf, Silicon Processing for the VLSI Era (1990).

$^{14}$ Y.-M. Lin, J. Appenzeller, and Ph. Avouris, Nano Lett.
\textbf{4}, 947 (2004).

$^{15}$ S. J. Wind, J. Appenzeller, R. Martel, V. Derycke, and Ph.
Avouris, Appl. Phys. Lett. \textbf{80}, 3817 (2002).\\
$^{1}$$^{6}$ A. Javey, J. Guo, D. B. Farmer, Q. Wang, D. Wang, R. G.
Gordon, M. Lundstrom, and H. Dai, Nano Lett. \textbf{4}, 447 (2004).

$^{17}$ Y.-M. Lin, J. Appenzeller, and Ph. Avouris, 62nd Device
Research Conference Digest 2004, p. 133.

\end{document}